\documentclass{svmult}

\usepackage{makeidx}         % allows index generation
\usepackage{graphicx}        % standard LaTeX graphics tool
\usepackage{multicol}        % used for the two-column index
\usepackage[bottom]{footmisc}% places footnotes at page bottom
\makeindex             % used for the subject index
\begin{document}
\newcommand{\be}{\begin{equation}}
\newcommand{\ee}{\end{equation}}
\newcommand{\lb}{\left(}
\newcommand{\rb}{\right)}
\newcommand{\eps}{\varepsilon}

\title*{Application of noise level estimation for  portfolio optimization}
\author{Krzysztof Urbanowicz\inst{1}\and
Janusz A. Ho{\l}yst\inst{2}}
\institute{Max Planck Institute for the Physics of Complex
Systems, N\"{o}thnitzer Str. 38, D--01187 Dresden, Germany
\texttt{urbanow@pks.mpg.de;http://www.chaosandnoise.org} \and
Faculty of Physics and Center of Excellence for Complex
Systems Research \\Warsaw University of Technology \\
Koszykowa 75, PL--00-662 Warsaw, Poland\\
\texttt{jholyst@if.pw.edu.pl}}
 \maketitle
\begin{abstract}
Time changes of noise level  at  Warsaw Stock Market are analyzed
using a recently developed method basing on properties of the
coarse grained entropy. The condition of the minimal noise level
is used to build an efficient portfolio. Our noise level approach
seems to be a much better tool for risk estimations than standard
volatility parameters. Implementation of a corresponding threshold
investment strategy gives positive returns for historical data.
\end{abstract}

{\it pacs 05.45.Tp,89.65.Gh} \par {\it keywords: Noise level
estimation, stock market data, time series, portfolio
diversification}
\section{Introduction}
\par Although it is a common believe that the stock
market behaviour is driven by  stochastic processes
\cite{voit,Buchound,Mantegna} it is difficult to separate
stochastic and deterministic  components   of market dynamics. In
fact the deterministic fraction  follows  usually from  nonlinear
effects and can possess  a non-periodic or even chaotic
characteristics \cite{Peters,Holyst}. The aim of this paper is to
study the level of stochasticity in time series coming from stock
market. We will show that our noise level analysis can be useful
for portfolio optimization.
\par We employ here a  method of noise-level estimation that has been
 described in details in \cite{urbanowicz}.
 The method is quite universal and it  is valid even for high noise
 levels.
The method makes use of a functional dependence of coarse-grained
correlation entropy $K_2(\eps)$ \cite{kantzschreiber} on the
threshold parameter $\eps$. Since the function $K_2(\eps)$ depends
in a characteristic way on the noise standard deviation $\sigma$
thus  $\sigma$ can be found  from a shape of  $K_2(\eps)$. The
validity of our method has been verified by applying it for the
noise level estimation in several chaotic models
\cite{kantzschreiber} and for the Chua electronic circuit
contaminated by noise. The method distinguishes a noise appearing
due to the presence of a stochastic process from a non-periodic
{\it deterministic} behaviour (including the deterministic chaos).
Analytic calculations justifying our method have been developed
for the gaussian noise added to the observed deterministic
variable. It has been also checked by numerical experiments  that
the method works properly for a uniform noise distribution and at
least for some models with a dynamical noise corresponding to the
Langevine equation \cite{urbanowicz}. The method has been already
successfully  applied for noise level calculations of engine
process \cite{urbengine} and has given  similar results to an
approach basing on neighboring distances in Takens space
\cite{urbtakens}.

\section{Choosing  low noise portfolio}

\par In the present paper we define the noise level as the ratio of standard deviation of
estimated noise $\sigma$ to the standard deviation of data
$\sigma_{data}$
\begin{equation} NTS=\frac{\sigma}{\sigma_{data}}\end{equation}
\par In the first step we construct a portfolio from $M$ stocks with the
minimal value of the stochastic variable \cite{urbAPFA4}. We
assume that one can do this by maximization of the following
quantity: \be \mathcal{B}=\sum\limits_{i=1}^{M}\sum\limits_{j=1}^M
p_i p_j
\frac{\sigma_{i,D}}{\sigma_{i}}\frac{\sigma_{j,D}}{\sigma_{j}}\rho_{i,j}=max
\end{equation} where $\sigma_{i,D}$ is the standard deviation of
deterministic part of the stock $i$, $\sigma_i$ is the standard
deviation of the noise for this stock and $\rho_{i,j}$ is the
correlation coefficient between deterministic parts  of stocks $i$
and $j$. The maximal value of $\mathcal{B}$ can be received  with
the help of the steepest descent method by changing  variables
$p_i$ and keeping the normalization constraint  $\sum_{i=1}^M p_i
= 1$.
\par In some cases for practical reasons it is more efficient not
to minimize the noise level in the portfolio but to maximize it.
This is because the method for noise level estimation can fail and
it can  occasionally give wrong values of $NTS$. When we minimize
the noise level it can happen that one stock with an artificially
very low noise level dominates the whole portfolio and the risk
increases without any additional profit.
\section{Investment method}
\par In our investment method we make use of additional information,
available due to the knowledge of the noise level, to increase
profits from selected portfolios. The simplest approach is to
introduce a threshold for a noise level. We divide all portfolios
into two classes: profitable and nonprofitable taking into account
high or low values of the noise level and a positive or a negative
past trend. The partition  into high/low noise classes is based on
the threshold parameter $NTS_{th}$ that should be optimized.
Additionally we label portfolio by calculations of an average
return for the last $N_{win}$ data. We use the following
algorithm: if the past trend from $N_{win}$ data of the portfolio
is positive $m_p>0$ and the noise level of the portfolio is small
($NTS_p<NTS_{th}$) we consider the portfolio as a profitable. We
have a  profitable portfolio also when it is more stochastic
($NTS_p>NTS_{th}$) but its trend is negative $m_p<0$. In the
remaining two cases we consider the portfolio as a nonprofitable.
In such a way we create the basic strategy giving $\{p_i\}$, which
involves the information on the noise level and the past trend in
the portfolio selection. This basic strategy should then be
adjusted using a {\it risk parameter} $\mathbf{r}$ that is
introduced below. The process of the final  portfolio selection is
based on the comparison of the optimized portfolio  to the
simplest portfolio consisting of equal contributions from all
stocks ($p_i=1/M$, $i=1,...,M$). We set up a composition of the
final portfolio $\{\tilde{p}_i\}$ with a use of certain risk
parameter $\mathbf{r}$ on the preliminary optimized portfolio
$\{p_i\}$ as follows: \be \tilde{p}_i=\frac{1}{M}+\mathbf{r}\lb
p_i-\frac{1}{M}\rb\label{eq3}\end{equation} One should mention
that for a negative value of the parameter $\mathbf{r}$ we have
the opposite investing to the composition $p_i$.
\par
\begin{figure}
  \hfill
  \begin{minipage}[t]{.45\textwidth}
\begin{center}
\includegraphics[scale=0.21,angle=-90]{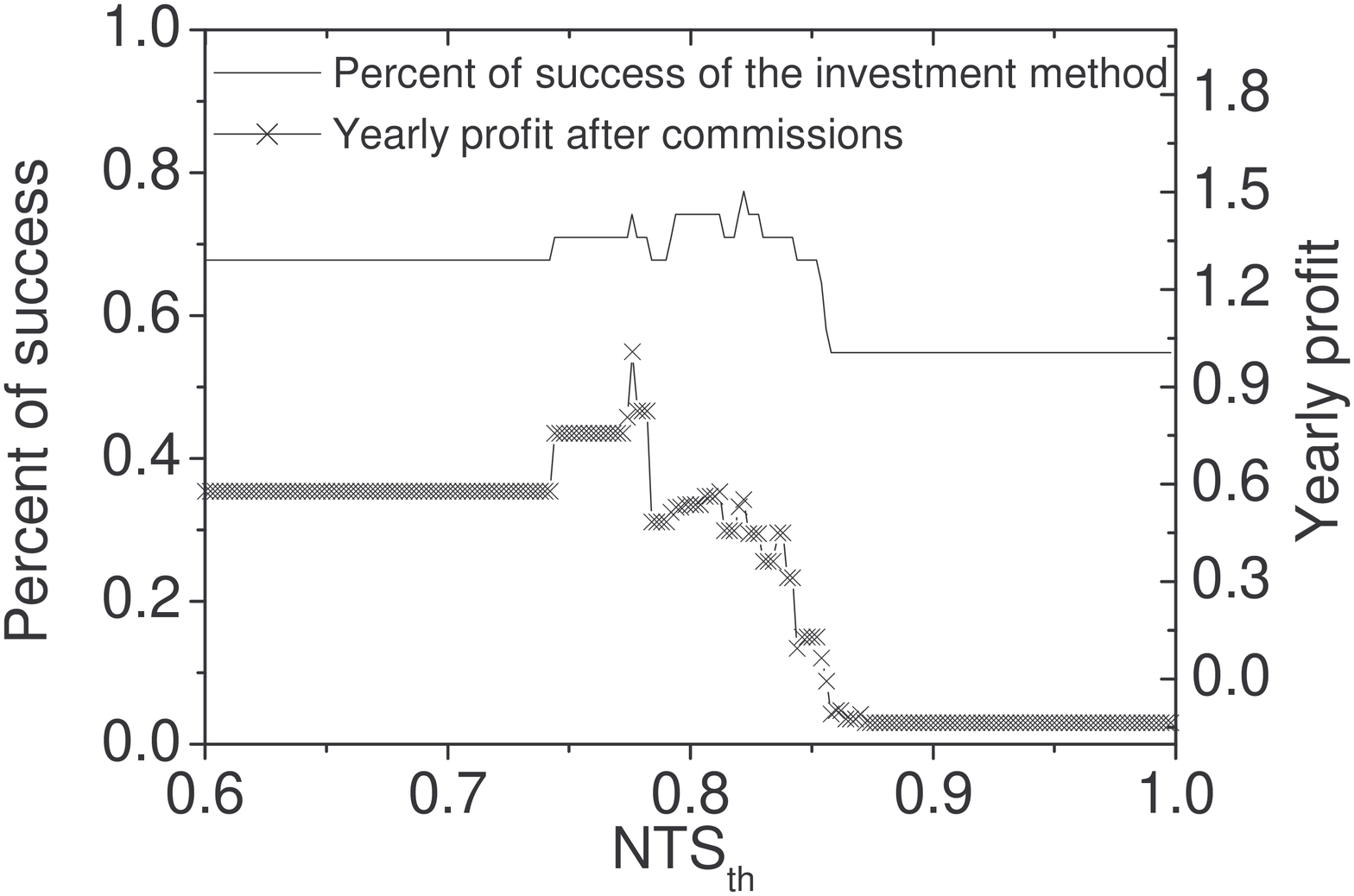}
\caption{\label{fig.NTSth} Plot of the investment success as a
function  of parameter $NTS_{th}$ in the period of January - July
2003 at the Warsaw Stock Exchange ($N_{win}=2500$). Portfolio
consists of 18 stocks from WSE.}
\end{center}
  \end{minipage}
  \hfill
  \begin{minipage}[t]{.45\textwidth}
    \begin{center}
\includegraphics[scale=0.21,angle=-90]{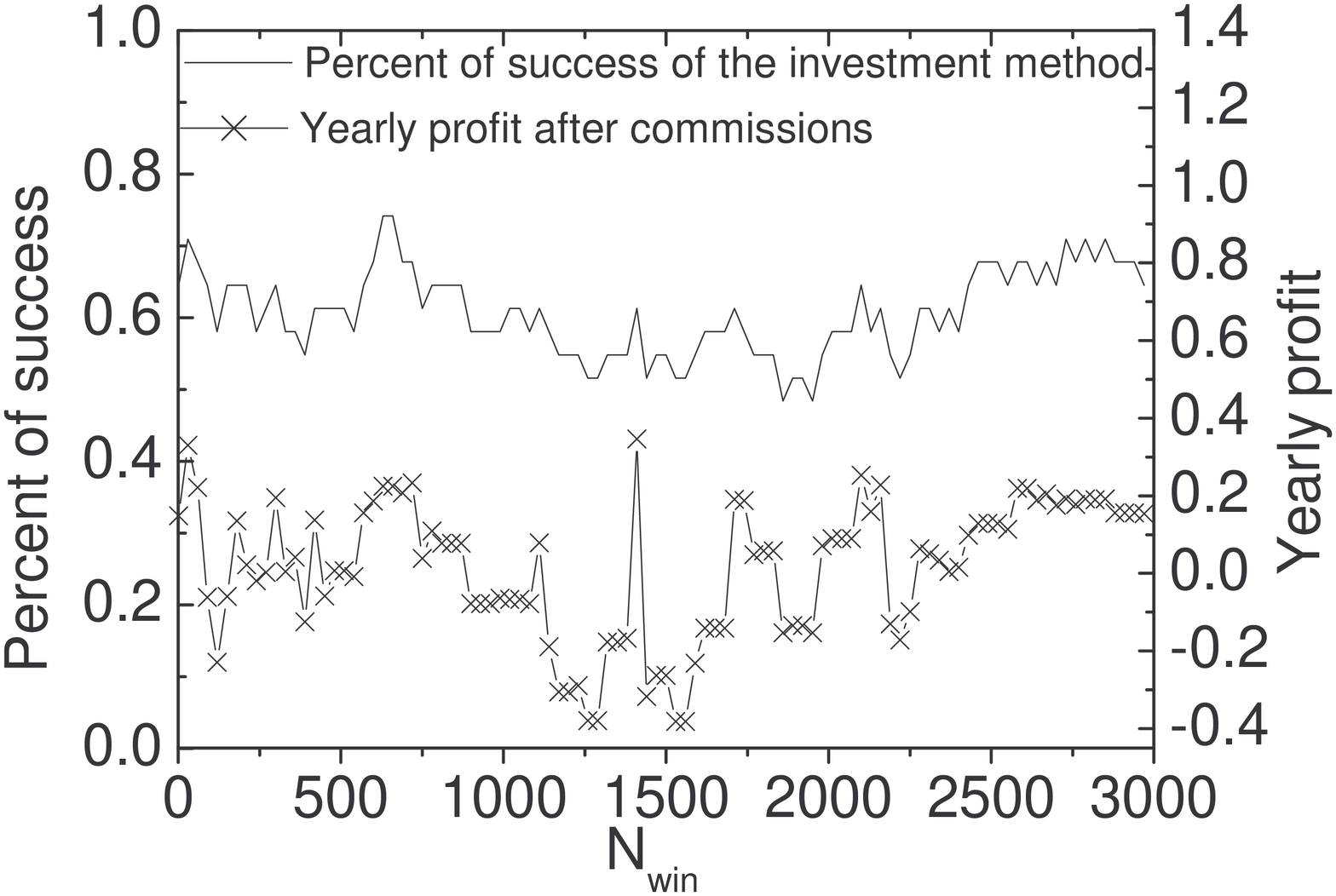}
\caption{\label{fig.Nwin} Plot of the investment success as a
function of parameter $N_{win}$ in the period of January - July
2003 at Warsaw Stock Exchange ($NTS_{th}=0.85$). Portfolio
consists of 18 stocks from WSE.}
    \end{center}
    \end{minipage}
\end{figure}

 At Fig.~\ref{fig.NTSth} the level of success of our
investment method as a function  of the parameter $NTS_{th}$ is
shown. Here the percent of success corresponds to a  fraction of
positive returns from our strategy. We have used a negative risk
parameter $\mathbf{r}=-10$ to get a positive profit for small
values of $NTS_{th}$: $NTS_{th}<0.85$ in the above simulations.
 A similar dependence on the parameter $N_{win}$ is shown at
Fig.~\ref{fig.Nwin}. The percent of the success in both cases is
above $50\%$ and for some regions of selected parameters the
strategy brings positive returns  after commissions deduction.
\par It is clear that  to use  our approach we have to find optimal threshold
parameters $NTS_{th}$ and $N_{win}$. Our optimization method is
quite straightforward and it resembles a genetic algorithm. During
the optimization process we change the selection probability for
actual  values of optimized parameters i.e. we increase the
probability if the profit from portfolio is positive and we
decrease in the opposite case. The optimization process is
terminated when we reach a satisfactory mean value of a yearly
profit from past data (here it is $30\%$). In such a way we
optimize simultaneously two parameters $N_{win}$ and $NTS_{th}$.
\par
\begin{figure}[ht]
  \hfill
  \begin{minipage}[t]{.45\textwidth}
    \begin{center}
\includegraphics[scale=0.22,angle=-90]{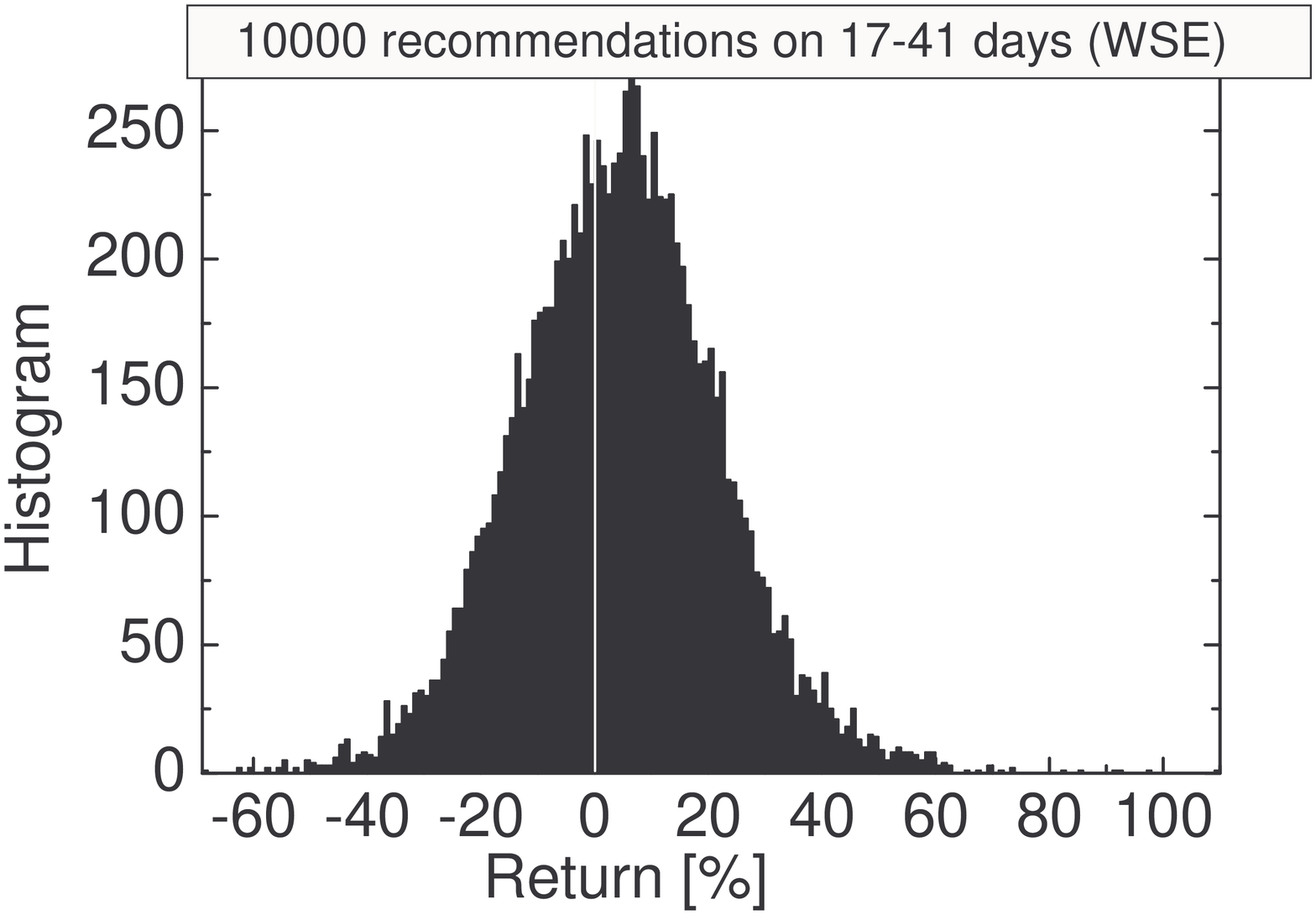}
\caption{\label{fig.distr} Histogram of returns received by our
strategy. The mean return equals to $4.33\%$ while the histogram
dispersion is about $17\%$.}

    \end{center}
  \end{minipage}
  \hfill
  \begin{minipage}[t]{.45\textwidth}
    \begin{center}
\includegraphics[scale=0.22,angle=-90]{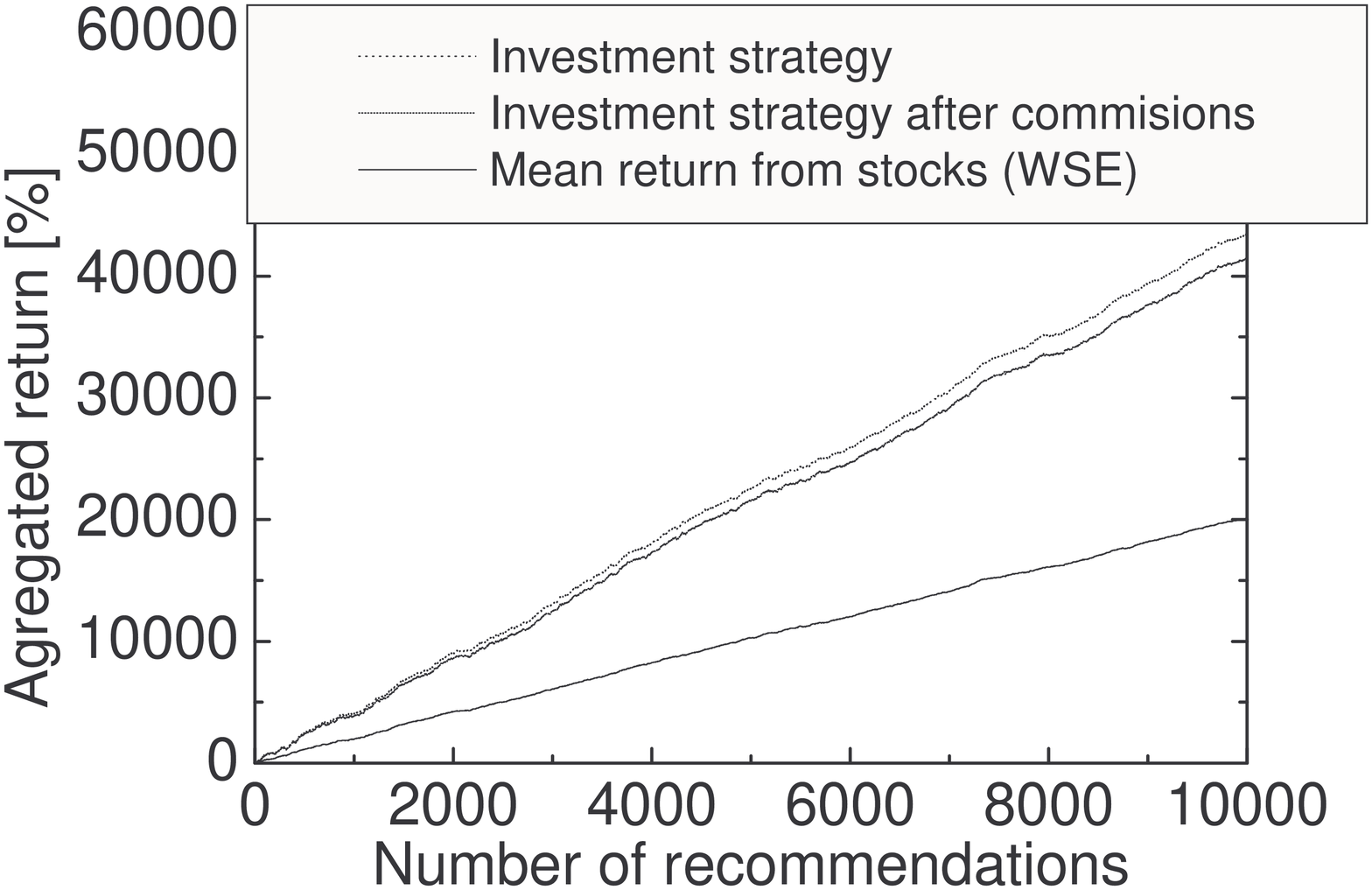}
\caption{\label{fig.returnWIG} The aggregated return for our
investment strategy applied for the Warsaw Stock Exchange. The
return corresponds to the mean annual return $56\%$ while the mean
annual return of Warsaw Stock Index was about $28\%$  at the same
time period.}
    \end{center}
  \end{minipage}
  \hfill
\end{figure}

 We begin our algorithm by generating randomly chosen stocks in the initial portfolio.
 Then we randomly select a    starting moment for our virtual investment.
 The next step  is to  optimize the parameters $N_{win}$ and $NTS_{th}$ using
available data from the period prior to the selected starting
point. Finally  we invest in the portfolio described by the risk
value $\mathbf{r}=-3$ (see Eq. \ref{eq3}). The procedure was
repeated $10000$ times and at the end we  calculated  an average
profit i.e. the efficiency of the method. At Fig.~\ref{fig.distr}
we show a distribution of returns for our portfolio at Warsaw
Stock Exchange. We have calculated recommendations for windows
$17-41$ days long on the period July 2002 - December 2003 (see
Fig.~\ref{fig.returnWIG}). The annual return received in such a
way after commissions substracting  is around $56\%$ (the
commission level has been set to $0.25\%$). To omit artificially
large price changes that can be caused by such effects as stock
splitting, extreme returns larger than $12$ standard deviation of
data have been rejected.
\section{Conclusions}
\par In conclusion we have analyzed noise level for data from Warsaw Stock Exchange.
We show that our noise level estimations can be useful for
portfolio optimization. The resulting investment strategy brings
larger profits than a simple average from the same stocks.

\printindex

\begin{thebibliography}{99.}
\bibitem{voit} Voit J (2001) The
Statistical Mechanics of Financial Markets. Springer-Verlag,
Berlin Heidelberg New York Barcelona Hong Kong London Milan Paris
Singapore Tokyo
\bibitem{Buchound} Bouchaud JP, Potters M (2000) Theory of financial risks - from
statistical physics to risk management. Cambridge University
Press, Cambridge
\bibitem{Mantegna} Mantegna RN, Stanley HE (2000) An
Introduction to Econophysics. Correlations and Complexity in
Finance. Cambridge University Press, Cambridge
\bibitem{Peters} Peters EE (1997) Chaos and Order in the
Capital Markets. A new view of cycle, Price, and Market
Volatility. John Wiley $\&$ Sons, New York.
\bibitem{Holyst} Ho{\l}yst JA et al. (2001) Observations of deterministic chaos in financial time series by recurrence plots, can one control chaotic economy? European Physical Journal B 20:531-535
\bibitem{urbanowicz} Urbanowicz K, Ho{\l}yst JA (2003) Noise-level estimation of time series using coarse-grained entropy. Phys. Rev. E 67:046218;
http://www.chaosandnoise.org
\bibitem{kantzschreiber} Kantz H, Schreiber T (1997) Nonlinear Time Series Analysis. Cambridge University Press, Cambridge
\bibitem{urbengine} Kaminski T et al. (2004) Combustion process in a spark ignition engine: Dynamics and noise level estimation. Chaos 14(2):461-466; Litak G et al. (2005) Estimation of a Noise Level Using Coarse-Grained Entropy of
Experimental Time Series of Internal Pressure in a Combustion
Engine. Chaos Solitons \& fractals 23(5):1695-1701

\bibitem{urbtakens} Urbanowicz K, Ho{\l}yst JA (2004) Noise estimation by the use of neighboring distance in Takens space and its application to the stock market data. Proceedings of the  Conference Complexity in science and society, International Journal of Bifurcation and Chaos, arXiv:cond-mat/0412098; http://www.chaosandnoise.org
\bibitem{urbAPFA4} Urbanowicz K and Ho{\l}yst JA (2004) Investment strategy due to the minimization of the noise level in a portfolio. Physica A 344:284-288
\end{thebibliography}
\end{document}